\documentclass[preprint,12pt,5p,twocolumn]{elsarticle}
\biboptions{sort&compress} 

\usepackage{amssymb}
\usepackage{amsmath, cuted} 
\AfterEndEnvironment{strip}{\leavevmode}
\usepackage{graphicx}
\usepackage{caption}
\usepackage{subcaption}
\usepackage{multirow}
\usepackage{multicol}
\usepackage{hyperref}
\usepackage{natbib}
\usepackage{url}
\usepackage{xcolor}
\usepackage{soul}
\usepackage{afterpage}
\usepackage{fancyhdr}

\journal{Computer Networks}

\fancypagestyle{firstpage}{%
 \fancyhf{}%
 \fancyhead[L]{\centering\tiny{This is the authors' version of the manuscript accepted in Computer Networks (2024).\\Final publication can be found in doi: https://doi.org/10.1016/j.comnet.2024.110569.\\This manuscript version is made available under the CC-BY-NC-ND 4.0 license https://creativecommons.org/license/by-nc-nd/4.0.}}
}

\begin{document}

\begin{frontmatter}



\title{A Spatio-temporal Prediction Methodology Based on Deep Learning and Real Wi-Fi Measurements}


\author[inst1]{Seyedeh Soheila Shaabanzadeh\corref{cor1}}
\cortext[cor1]{Corresponding author}

\affiliation[inst1]{organization={Signal Theory and Communications},
            addressline={Universitat Politècnica de Catalunya (UPC), Carrer de Jordi Girona, 1-3}, 
            postcode={08034}, 
            state={Barcelona},
            country={Spain}}

\author[inst1]{Juan Sánchez-González}


\begin{abstract}
The rapid development of Wi-Fi technologies in recent years has caused a significant increase in the traffic usage. Hence, knowledge obtained from Wi-Fi network measurements can be helpful for a more efficient network management. In this paper, we propose a methodology to predict future values of some specific network metrics (e.g. traffic load, transmission failures, etc.). These predictions may be useful for improving the network performance. After data collection and preprocessing, the correlation between each target access point (AP) and its neighbouring APs is estimated. According to these correlations, either an only-temporal or a spatio-temporal based prediction is done. To evaluate the proposed methodology, real measurements are collected from 100 APs deployed in different university buildings for 3 months. Deep Learning (DL) methods (i.e. Convolutional Neural Network (CNN), Simple Recurrent Neural Network (SRNN), Gated Recurrent Unit (GRU), Long Short-Term Memory (LSTM), Transformer) are evaluated and compared for both temporal and spatio-temporal based predictions. Moreover, a hybrid prediction methodology is proposed using a spatial processing based on CNN and a temporal prediction based on RNN. The proposed hybrid methodology provides an improvement in the prediction accuracy at expenses of a slight increase in the Training Computational Time (TCT) and negligible in Prediction Computational Time (PCT).
\end{abstract}




\end{frontmatter}


\thispagestyle{firstpage}  

\section{Introduction}
\label{sec:introduction}
In the last years, we have witnessed a remarkable growth in the amount of user traffic in wireless networks due to the introduction of new multimedia services such as high-quality video, augmented/virtual reality, etc. These new services have very strict requirements in terms of reliability, latency, bit error rate, etc. \cite{cisco2020cisco}. A solution to deal with these requirements and traffic demands is the deployment of small cells operating cellular technologies (e.g. 4G, 5G) combined with the deployment of Wi-Fi technologies using the unlicensed band in certain hotspots \cite{gacanin2017wi, kibria2018big}. 

The recent development of (Big) Data monitoring and analytics technologies is one of the main pillars in cellular/Wi-Fi networks to cope with the previously mentioned challenges. In particular, the capacity to collect information about the users and the network by deploying monitoring tools, and the capacity to extract knowledge of the collected data by means of Artificial Intelligence/Machine Learning (AI/ML) technologies can help the network to be smarter, more intelligent, self-aware, cost-efficient, and self-optimized \cite{kibria2018big}. 

The knowledge obtained by means of prediction, based on e.g. regression or classification algorithms, can be useful for supporting different decision-making processes over the network related to network reconfiguration and optimization. In particular, knowledge about future network traffic can be useful for different purposes such as proactive resource allocation, prevention of congestion/overload situations or activation of load balancing processes to distribute network traffic among multiple cells/APs. Several approaches have been proposed in this context for cellular \cite{zhang2019new,wang2017spatiotemporal} and Wi-Fi \cite{thapaliya2018predicting, khan2020real, chen2021flag, sone2020wireless, sone2021forecasting} networks. 

In the context of traffic prediction/classification, multiple approaches can be found in the literature. For instance, papers such as \cite{aceto2019mobile,aceto2019mimetic}, proposed the use of Deep Learning (DL) techniques for mobile traffic classification. On the other hand, for traffic prediction, other works such as, \cite{deng2021short,xu2017high,nie2017network,li2020prophet} make use of a Gaussian Model, while \cite{wei2012intrusion} creates a data traffic prediction model based on autoregressive moving average (ARMA). 
Other approaches consider both temporal and spatial traffic analysis to improve the prediction accuracy. Spatio-temporal prediction combines spatial dependencies and interactions, capturing how traffic at one location impacts neighbouring areas. It considers traffic propagation effects, handles spatial variability, and adapts to dynamic network conditions. By modeling spatial and temporal features together, it provides contextually aware, accurate, and adaptive predictions for network metrics like traffic load and failures, optimizing network management and resource allocation. Several works proposed a joint space/time traffic prediction, assuming that the traffic is temporally and spatially correlated \cite{wang2017spatiotemporal,zhang2018citywide,huang2017study,li2020deep, wen2020assisting}. In particular, in the context of cellular networks, a spatio-temporal prediction has been presented with CNN-RNN \cite{huang2017study,li2020deep}, by combining auto-encoder and LSTM \cite{wang2017spatiotemporal}, by CNN-based framework \cite{zhang2018citywide}. In recent years, other novel approaches have been proposed for prediction. As an example, transformer-based deep learning models have been proposed to tackle the task of traffic time series prediction \cite{li2022lightweight, hu2022citywide, gu2023spatial, yimeng2022prediction}.

To the best of the authors’ knowledge, most of the papers that can be found in the literature focus on traffic prediction in cellular networks. However, very few papers deal with traffic prediction in Wi-Fi networks. In this context, the authors of \cite{thapaliya2018predicting} predict different congestion levels (low, medium or high) in different locations in a University Campus by means of a support vector regression (SVR) to predict network features and then, expectation method (EM) to obtain different levels of congestion.  
In \cite{khan2020real}, data-driven ML techniques are used to predict the transmission throughput in Wi-Fi networks. Moreover, \cite{chen2021flag} extracted three categories of features, i.e., individual features, spatial features, and temporal features from network data, that can be used to label each AP. Then, with these features, a DL architecture was implemented, which consists of two separated deep RNN models. Finally, based on the training model and the input of historical client traffic load, the future client traffic loads are predicted in an online process. In \cite{sone2021forecasting}, a temporal prediction of channel and traffic utilization was performed in order to better allocate the resources in a proactive manner. 

Regarding this topic, paper \cite{sone2020wireless} is the only one that is directly close to our work (i.e., spatio-temporal prediction in Wi-Fi systems). The paper \cite{sone2020wireless} highlights the importance of incorporating measurements from highly correlated AP neighbours to improve prediction accuracy. It analyzes data from 8 APs, while our study expands this analysis to include more APs. Moreover, this paper incorporates more recent prediction methods, such as Transformers, in addition to other DL techniques proposed in \cite{sone2020wireless}. Unlike the approach of \cite{sone2020wireless}, which provides a single time complexity for both training and prediction, our paper separately calculates Training Computational Time (TCT) and Prediction Computational Time (PCT) that can effectively aim to address implementation aspects, challenges and considerations for real-world deployment. 
Within this context, the main novelty with respect to previous works is the proposal of a general framework for the prediction of future values of a specific network metric for a given AP in Wi-Fi network.  
In particular, the main contributions of this paper are the following: 
\begin{itemize}
    \item The paper addresses spatio-temporal prediction of a specific Wi-Fi metric by developing an adaptive and context-aware prediction framework. This framework dynamically adjusts to spatial dependencies among APs, enhancing the accuracy and reliability of traffic predictions. By intelligently selecting whether to include information from neighbouring APs based on identified spatial correlations, our methodology overcomes the challenge of incorporating spatial dynamics into Wi-Fi metric prediction.
    \item In order to achieve a more comprehensive and data-driven solution to Wi-Fi metric prediction, a wide range of techniques including more recent prediction techniques (i.e. Transformers), alongside single and hybrid NN techniques are compared in terms of prediction accuracy and computational time.
    \item The proposed methodology is particularized for the prediction of future values of traffic at the AP and the prediction of transmission failures, but the methodology could be easily extended for other metrics.
    \item The proposed methodology is evaluated in a real Wi-Fi network with measurements collected from a large number of APs (i.e. 100 APs) over three months. This addresses the challenge of scalability in designing prediction systems for Wi-Fi networks, ensuring that the methodology can effectively handle larger network environments without compromising performance.
\end{itemize}

The rest of the paper is organized as follows. Section \ref{secmethod} explains the details of the proposed prediction methodology. Section \ref{secpretechniques} describes the techniques used for prediction. The scenario description presented in Section \ref{secscenario}, and the obtained results are discussed in Section \ref{secresults}. Section \ref{secImpleAspects} explains the implementation aspects necessary for deploying the methodology in a real network. Finally, conclusions are drawn in Section \ref{secconclusion}.

\section{Proposed Methodology Description}
\label{secmethod}
\begin{figure*}[t]
  \centering
      \includegraphics[width=1\linewidth]{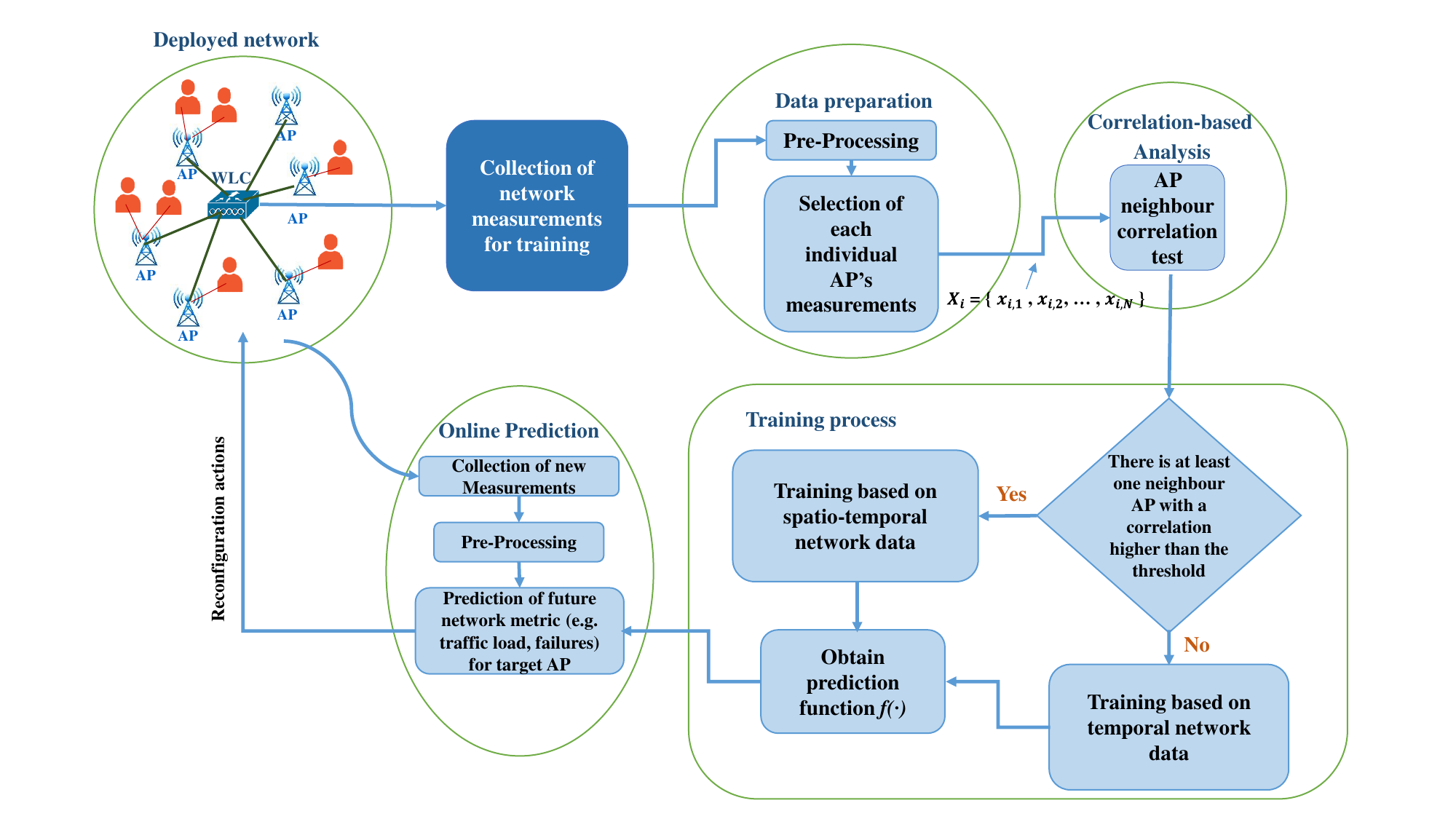}  
      \caption{General architecture of proposed methodology.}
  \label{fig:c3figure1}
\end{figure*}

This section presents the proposed prediction methodology and its applicability in the context of a Wi-Fi network. As shown in Fig. \ref{fig:c3figure1}, network measurements collected at the APs are sent to the Wireless LAN Controller (WLC) and finally stored in a centralized database located in the Data Management System that contains the historical measurements of the different APs. Then, with this data, an offline training is run in order to extract knowledge from the collected data. This offline training is done with certain periodicity in order to keep updated the obtained model according to recent collected measurements. Then, the prediction phase is done online and consists on making use of new measurements collected at the APs together with the models that were obtained in the training phase.

The obtained results in the prediction process may be useful for making adequate reconfiguration actions over the Wi-Fi network to improve the network performance \cite{9363693,kumar2008wireless}. For example, the prediction of a future peak of traffic in the area covered by an AP may be useful for e.g. the reconfiguration of the admission control threshold in order to control the amount of traffic at the AP or the activation of load balancing processes to encourage users to connect to less overloaded APs/radios.

The proposed methodology, assumes a Wi-Fi Network consisting of $I$ APs. Measurements are collected from the different $I$ APs with a periodicity of $T$ over a total of $D$ days. The time sequence of a specific collected metric of the $i$-th AP is represented as $X_i = \{x_{i,1}, x_{i,2} ,..., x_{i,N}\}$ where $i = 1, 2,…, I$, and $N$ is the total number of samples for each AP collected during the $D$ days. With this data, the proposed methodology aims to predict future values of this specific metric in a given AP by using the last $L$ collected measurements. The methodology can be used for the prediction of different possible metrics (e.g. related to network traffic, network failures, network performance, etc.). 	

After data collection, in the data preparation phase, the methodology runs a data pre-processing according to the following steps: (i) The first step consists on automatically determining missing data in some time periods. This may happen due to e.g. possible errors in the collection of measurements, etc. Missing data can be completed according to different possible techniques. In this paper, backfilling \cite{brownlee2018deep} is used, in which missing values at certain time period are filled with the data of the previous time period. If there are more than two consecutive time periods with missing data, the measurements of this day are removed and are not considered for training. (ii) The methodology determines the optimum value of the $L$. This process is done by the Augmented Dickey-Fuller Test that uses an autoregressive model for determining the most adequate value of $L$ \cite{brownlee2018deep}. The Augmented Dickey-Fuller test iteratively assesses the autocorrelation structure of time series data for various lag values. It selects the lag value that minimizes the test statistic, indicative of optimal lag for the time series, to evaluate its stationarity. For each AP, the optimum value of $L$ is obtained. Then, the most frequent value is selected. (iii) The input data $X_i$ is normalized so that all values are within the range between 0 and 1. 

After the pre-processing step, the proposed methodology calculates the correlation between each AP and all its neighbours. AP neighbours are determined by means of the Neighbour Discovery Protocol (NDP) \cite{cisco2016}. The criteria for the finding neighbour APs is ``received power". According to this, each AP measures the received power from nearby APs and determines a list of neighbours automatically with the APs detected with a received power higher than a specific threshold. In particular, the correlation between the $i$-th and the $j$-th APs (with $i \neq j$) is calculated according to (\ref{eq:c3equation1}):

\begin{strip}
\begin{equation}
\label{eq:c3equation1}
C_{i,j} = \frac{N(\sum_{n=1}^{N}x_{i,n}x_{j,n}) - (\sum_{n=1}^{N}x_{i,n})(\sum_{n=1}^{N}x_{j,n})}{\sqrt{[N\sum_{n=1}^{N}x_{i,n}^2-(\sum_{n=1}^{N}x_{i,n})^2][N\sum_{n=1}^{N}x_{j,n}^2-(\sum_{n=1}^{N}x_{j,n})^2]}}\ 
\end{equation}
\end{strip}
\noindent where $x_{i,n}$ and $x_{j,n}$ are the $n$-th measurement of the selected $i$-th AP and $j$-th AP, respectively. These correlations $C_{i,j}$ are used to determine the list of highly correlated neighbours of each AP. Then, if the correlation $C_{i,j}$ is higher than a specific threshold (i.e. $C_{i,j}>Ths$), the $j$-th AP is included in the list of highly correlated neighbours of the $i$-th AP. Then, for each $i$-th AP, a list of $M_i$ highly correlated neighbours is determined. In case that no highly correlated neighbour has been found for a given $i$-th AP (i.e. $M_i=0$), then, the proposed methodology makes a temporal based prediction using the $L$ previous measurements of this AP in order to predict the future value of this specific metric. Otherwise, (i.e. $M_i>0$), a spatio-temporal prediction is done based on the $L$ previous measurements of the $i$-th AP and also the $L$ previous measurements of its $M_i$ highly correlated neighbours.

With this data, an offline training is done in order to obtain a $f(.)$ function that will be used in the prediction step, as detailed hereinafter. For the case of the only temporal-based prediction, this $f(.)$ function is determined by observing the relationship between a specific measurement and its previous consecutive $L$ measurements for a give AP. Specifically, the prediction function $f(.)$ is determined according to the following. The set of measurements of the $i$-th AP $X_i = \{x_{i,1},  ..., x_{i,N}\}$ is split into $P$ different training tuples $X_i^P = \{x_{i,p},..., x_{i,p+1},..., x_{i,p+L}, x_{i,p+L+1}\}$ with $p=1,...,N-L-1$. Therefore, the $f(.)$ function is learnt by a DL process that consists on observing the relationship between $x_{i,p+L+1}$ and its $L$ previous elements $= \{x_{i,p},..., x_{i,p+1},..., x_{i,p+L}\}$ for all the $P=N-L-1$ tuples. The rationale of this approach is to identify patterns for target AP’s time series values of a network metric in $L$ consecutive time periods. 

\begin{figure}[h]
     \centering
         \includegraphics[width=1\linewidth]{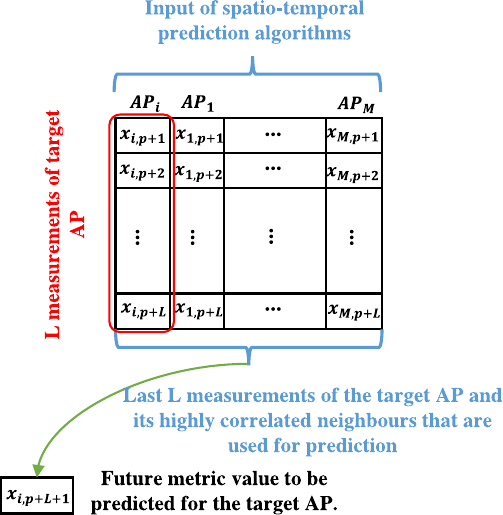} 
         \caption{Input and output data in the spatio-temporal prediction process.}
     \label{fig:c3figure2}
 \end{figure}

In turn, for the case of space-time-based prediction, not only the $L$ previous measurements of the corresponding $i$-th AP but also the $L$ previous measurements of each of the $M_i$ highly correlated neighbours are used for the prediction of the next sample of $i$-th AP (i.e. the measurements $\{X_i^P\}$ and the measurements of the $M_i$ highly correlated neighbouring APs $\{X_m^P\}$ with $m = 1,…, M_i$), for the prediction of $x_{i,p+L+1}$, as illustrated in Fig. \ref{fig:c3figure2}.

In the online prediction for a specific $i$-th AP, (see Fig. \ref{fig:c3figure1}), the $L$ measurements of this $i$-th AP and its $M_i$ highly correlated neighbours are used as input. Then, this input data is pre-processed in a manner similar to that in the training step. Finally, the prediction step makes use of the input pre-processed data and the learnt $f(.)$ function in order to make the prediction in this specific AP.

\section{Prediction Techniques}
\label{secpretechniques}
In this paper, we perform both temporal prediction and spatio-temporal prediction. In temporal prediction, the input is a vector with a length of L. In spatio-temporal prediction, the input is a matrix with a size of L*(M+1). Since M may be different for each AP and time series techniques typically require fixed-size input data, we utilize padding/truncation method \cite{brownlee2017long} to ensure a consistent input size before employing single or hybrid algorithms, which is a common necessity across these techniques. For practical reasons, we establish a pre-determined maximum value for M, indicating that a certain number of neighbouring APs is sufficient for accurate predictions. If the actual number of highly correlated neighbour APs is less than the maximum value of M, the algorithm pads the input data with zeros for missing features corresponding to non-existent neighbours. On the other hand, if the number of highly correlated neighbour APs exceeds the maximum value of M, the algorithm selects the M APs with the highest correlation and discards the rest. Prior to the rise of DL techniques, traditional or classical methods (e.g. Autoregressive Moving Average (ARMA)/Autoregressive Integrated Moving Average (ARIMA), Exponential Smoothing (ETS)), were commonly employed for time series analysis and prediction \cite{woodward2022time}. 
Deep Learning (DL) algorithms are suitable for identifying complex patterns that cannot be captured using conventional statistical ML. Due to their great performance, they have been used for different purposes such as time-series data prediction \cite{woodward2022time,zhang2019toward}. 
As \cite{sone2020wireless,huang2017study} suggest, DL algorithms within neural networks (NN) \cite{yegnanarayana2009artificial} such as LSTM, GRU, CNN, and their combinations are suitable for modeling and forecasting time series data related to wireless traffic usage. In recent years, new predictors (e.g. based on Transformers \cite{vaswani2017attention}) have been proposed and developed. Transformer, originally designed for natural language processing \cite{vaswani2017attention}, has recently become an important interesting tool for time series prediction. It includes a so-called attention mechanism that can be useful for capturing dependencies between different time steps in a sequence \cite{wu2020deep}.  
A brief description of the different algorithms considered in this paper are presented below: 
\begin{itemize}
    \item \textbf{ARIMA Algorithm:} AutoRegressive Integrated Moving Average (ARIMA) algorithm is a popular time series prediction method \cite{woodward2022time}.In our study, we apply ARIMA to measurements of the i-th target AP using a two-step process. First, we determine the appropriate model order using the auto-arima function. Once the optimal order is identified, we use ARIMA with that specific order along with the "Recursive Sampling" technique \cite{shumway2000time} to improve prediction performance. Recursive Sampling involves iteratively applying the prediction model, updating it with each new prediction for the next forecast. This process continues, making one-step-ahead forecasts and incorporating each predicted value into the dataset, until the desired forecast horizon is reached. This method is utilized only for temporal prediction. 
    
    \item \textbf{RNN Algorithms:} Recurrent Neural Networks (RNN) are a class of models effective for time-series prediction due to their ability to recurrently utilize previously processed data, improving prediction accuracy. SimpleRNN (SRNN) was one of the initial approaches, incorporating both current and past information \cite{talathi2015improving}. However, SRNN faces the vanishing gradient problem, limiting its ability to retain long-term information. To address this, Long Short-Term Memory (LSTM) was introduced, enabling better control over data flow and memory retention \cite{brownlee2018deep,kong2017short}. Despite its benefits, LSTM requires more computational resources and time due to its complexity. Gated Recurrent Unit (GRU), a variant of LSTM, reduces computational complexity but may sacrifice some long-term memory capacity \cite{ravanelli2018light}. 
    A comparison of the RNN algorithms is performed for both temporal and spatio-temporal prediction. 
    
    \item \textbf{CNN Algorithm:} Convolutional Neural Network (CNN) is also a type of NN which is not originally designed for modeling time-series data. However, several approaches \cite{brownlee2018deep}\cite{borovykh2017conditional} have adapted CNN algorithms for time series data processing leading to good performance results.  A possible way to adapt the input data for CNN algorithm is to divide the sequence into multiple input/output patterns, where a certain number of time steps are used as input and a time step is used as output to predict the next step. CNN utilizes three layers - filter, pooling, and fully connected - combined with the sequence to generate the output. CNN is used for both temporal and spatio-temporal prediction. 
     
    \item \textbf{CNN-RNN hybrid algorithm:} This architecture, only used for spatio-temporal prediction, combines CNN and RNN to leverage CNN's spatial feature learning and RNN's temporal dependency capture \cite{huang2017study}, as seen in previous applications such as activity recognition and video description \cite{donahue2015long}. 
    Fig. \ref{fig:c3figure3} illustrates the proposed architecture that combines CNN and RNN.
    As depicted in Fig. \ref{fig:c3figure3}, the CNN architecture integrates the temporal measurements of the $i$-th AP and its $M$ highly correlated APs to analyse spatial dependencies. Following two convolutional layers with 1D filters and two pooling layers, the dimensionality of the feature set is decreased. For instance, in the initial input vector $\{x_{i,p+1}, x_{1,p+1},..., x_{M,p+1}\}$, the resulting CNN output becomes $\{y_{1,p+1},..., y_{M-2,p+1}\}$. Here, $x_{i,p+1}$ denotes the target AP, while $\{x_{1,p+1},..., x_{M,p+1}\}$ represents the highly correlated APs, totaling M+1 features for input and M-2 features for output. The CNN output serves as input for the subsequent RNN phase (see Fig. \ref{fig:c3figure3}) to predict the metric's value in the subsequent time period ($x_{i,p+L+1}$).

    \begin{figure}[t]
     \centering
         \includegraphics[width=1\linewidth]{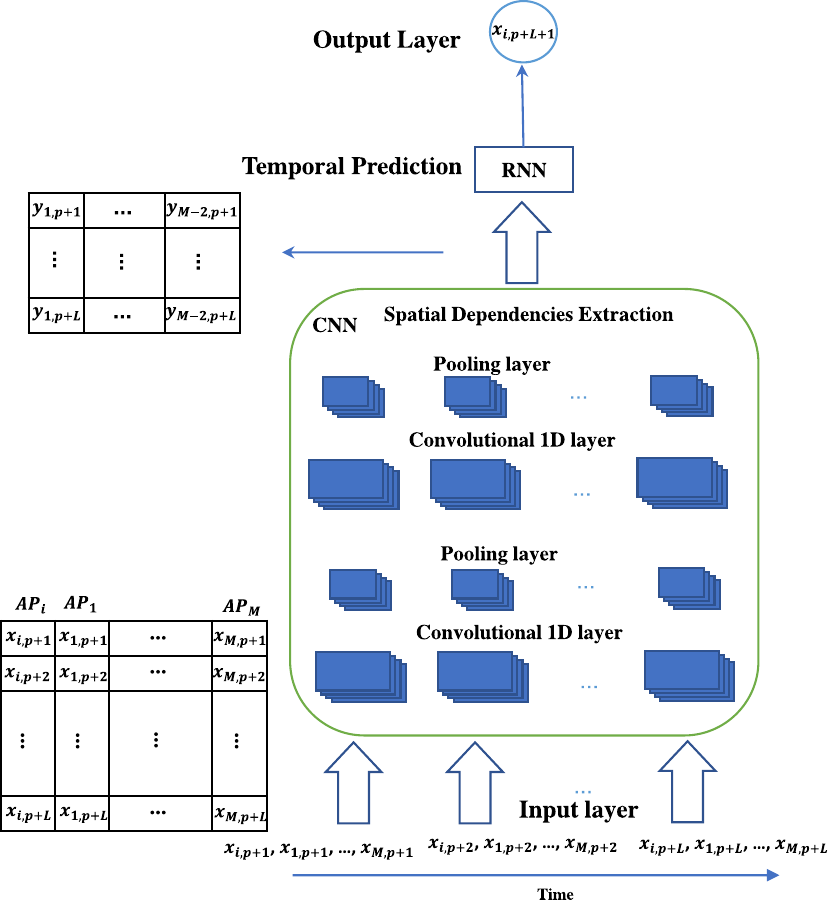}
         \caption{Hybrid CNN-RNN Architecture.}
     \label{fig:c3figure3}
    \end{figure}

    \begin{figure}[t]
      \centering
         \includegraphics[width=1\linewidth]{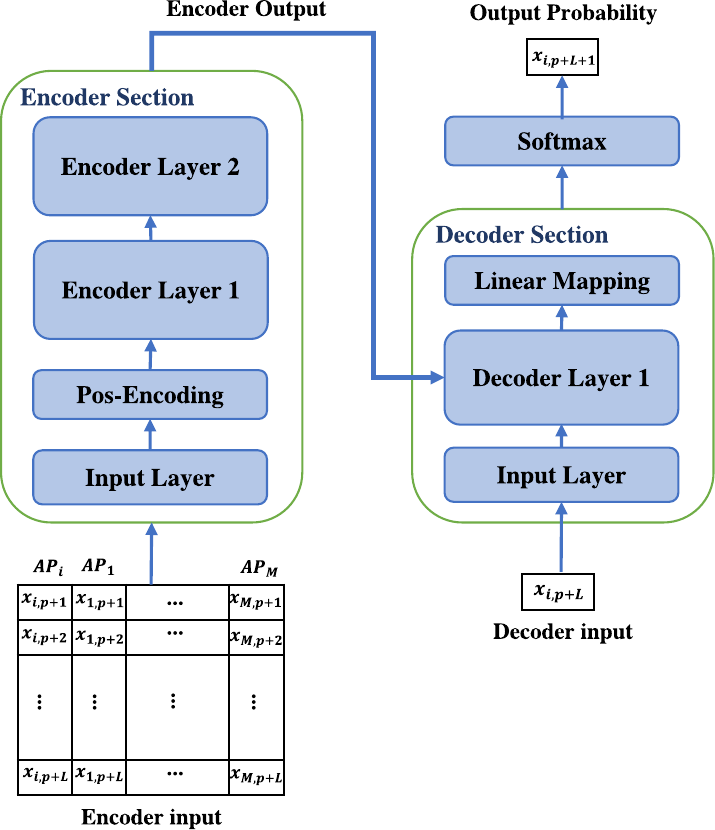}
         \caption{Transformer Architecture.}
      \label{fig:c3figure4}
    \end{figure}
    
    \item \textbf{Transformer algorithm:} This model is utilized for both temporal and spatio-temporal prediction. The architecture of the model consists of an encoder and a decoder stack \cite{vaswani2017attention}. 
    Fig. \ref{fig:c3figure4} depicts the general architecture for spatio-temporal prediction based on Transformer methodology. The available data is used as input to the encoder while the decoder operates on the sequence input of $x_{i,p+L}$ to produce the decoder output $x_{i,p+L+1}$. Throughout the encoding phase, the traditional canonical self-attention mechanism is substituted with the ProbSparse self-attention as implemented in \cite{zhou2021informer}. All the other sub-layers of the encoder remain unchanged and follow the standard configuration of a normal encoder. In ProbSparse attention, each key is permitted to attend exclusively to the dominant queries, a distinctive characteristic that enhances the model’s efficiency and performance in handling spatio-temporal predictions. Concerning the decoder, a standard decoder structure based on \cite{vaswani2017attention} is adopted. The decoder is composed of a stack of two identical multi-head attention layers.
      
\end{itemize}

\section{Experimental Design}
\label{secscenario}
For the evaluation of the proposed methodology, real measurements were collected from a total of 100 APs deployed in a University Campus of Universitat Politécnica de Catalunya, located in Barcelona during three months. The collection and management of the measurements is done by means of the Cisco Prime Infrastructure \cite{infrastructure3}. This tool periodically collects a large amount of AP measurements to capture the network status and centralizes all this information for building statistics useful for prediction purposes, etc. It is worth mentioning that the proposed prediction approach takes advantage of the information available at the Cisco Prime Infrastructure, without requiring any extra information and without generating any extra communication overhead. 

The proposed prediction methodology can be useful for the prediction of different user metrics. As an example, the prediction methodology is used here for the prediction of future AP traffic values and transmission failures. We preprocessed the data in a form of time series data for each AP, separately. The prediction methodology makes use of the measurements collected during the weekdays between September 9th, 2019 and December 22th, 2019 from 5:30 to 22:00 for every day with a periodicity $T$=30 minutes. This corresponds to 34 time periods per day, resulting in a total of D=75 days and N=2550 measurements for each metric.

After pre-processing, the measurements of traffic load and transmission failures for each single AP are obtained. These two metrics are defined in equations (\ref{eq:c3equation2}) and (\ref{eq:c3equation3}). According to (\ref{eq:c3equation2}) the normalized traffic is calculated as the number of transmitted packets in each time period ($TxC$) normalized by the maximum observed value in the measurements data ($MaxV$). On the other hand, the traffic transmission failures are calculated according to equation (\ref{eq:c3equation3}) where, $FTx$, $STx$ and $STxR$ represent the number of failed transmitted packets, the number of successfully transmitted packets and the number of successfully transmitted packets after retransmissions, respectively. 

\begin{equation}
\text{Traffic Load} = \frac{TxC}{MaxV} 
\label{eq:c3equation2}
\end{equation}
\begin{equation}
\text{Failures} = \frac{FTx}{STx+STxR} 
\label{eq:c3equation3}
\end{equation}

\begin{figure}[t]
     \centering
     \includegraphics[width=1\linewidth]{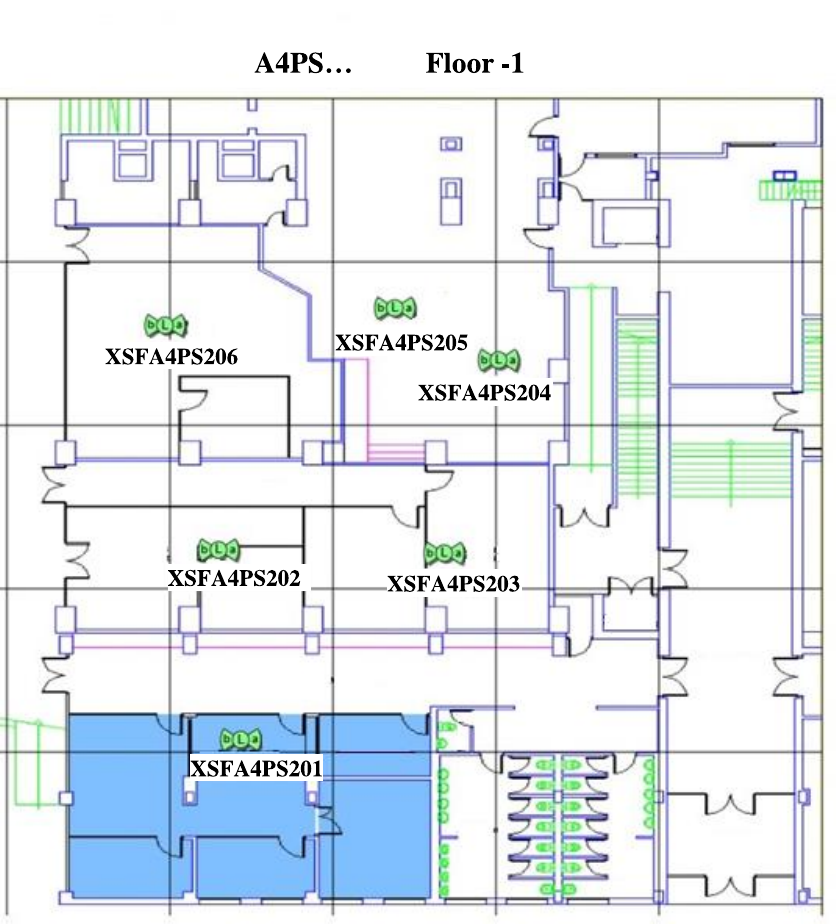}
     \caption{Location of APs in study and class area.}
     \label{fig:c3figure5}
\end{figure}

In order to illustrate the methodology, first, a small scenario consisting on 6 APs is considered as shown in Fig. \ref{fig:c3figure5}. In this small scenario, for temporal prediction, we aim to predict future values of traffic and failures in a specific AP (i.e. XSFA4PS205). Then, the analysis is extended for 100 APs located in seven different buildings, each with three floors, in order to validate the obtained results.

The collected dataset includes numerous features, but we specifically utilize the following: 
\begin{itemize}
    \item Time: this corresponds to the timestamp when the measurements have been taken.
    \item AP\_name: this feature is related to the name of AP.
    \item Radio\_type: it indicates the radio interface where the measurements are taken, in our case, predicting traffic in the 2.4GHz band.
    \item Number of transmitted packets (TxC): this metric is a counter that is increased every time a MSDU (MAC Service Data Unit) is transmitted. 
    \item Number of successfully transmitted packets (STx): this metric is a counter that is increased every time a MSDU is transmitted successfully without the need of retransmission.
    \item Number of successfully transmitted packets after retransmissions (STxR): this is a counter that is increased every time a MSDU is successfully transmitted after one or more retries. 
    \item Number of failed transmitted packets (FTx): this metric is a counter that is increased every time a MSDU frame is not transmitted successfully after a maximum number of retransmission attempts. 
\end{itemize}
These counters are collected with a given periodicity (T=30 minutes) and set again to zero. Each time period in the dataset contains values for all features. The ``AP\_name" and ``Radio\_Type" values are strings, ``Time" is in date/time format, and the remaining features are integers. To prepare the dataset for prediction, we perform several preprocessing steps. Initially, traffic load and failure measurements for each AP are generated using equations \ref{eq:c3equation2} and \ref{eq:c3equation3}, respectively (i.e. after this calculation we have 2 features of traffic load and failures for each AP). These measurements are then normalized by dividing them by the maximum collected measurement. Subsequently, they are converted to measurements with a specific periodicity of T=30 using the ``asfreq" technique in the Pandas Python library. During this conversion, any missing data is filled using the backfilling technique, ensuring that no more than one subsequent missing data point is encountered. Following these preprocessing steps, a total of 2550 observations are available for each AP, ready for further analysis and prediction. Before beginning the training phase for each AP, we selected M=5 as the maximum number of neighbours because considering the five most correlated neighbour APs captures sufficient spatial dependencies for accurate predictions without overwhelming the model with excessive information. In addition, we utilize the Augmented Dickey-Fuller test to determine the optimal lag. When conducting temporal prediction with only one feature and one column, the test straightforwardly identifies the lag. However, in the spatio-temporal prediction, where we consider multiple correlated APs (with column 1 representing the main AP measurement and additional columns representing highly correlated APs), we select the maximum lag value among all columns. For example, to compute lag for the target AP of XSFA4PS205 and its neighbours, the lag for XSFA4PS201, XSFA4PS203, XSFA4PS204, XSFA4PS205, XSFA4PS206 are 25, 27, 25, 25, 25. And the max of lag among them is 27. This approach ensures that we effectively capture the temporal dependencies present in the data. However, we have observed that the lag values for each column are often very close, resulting in minimal impact on prediction performance. 

Following the dataset preprocessing for each RNN, CNN, and Hybrid model implementation, we use a validation method called walk-forward validation (also known as expanding window validation), suitable for time series data \cite{brownlee2017introduction} with a 0.3 percentage split. 
The model is trained by using past data and is validated on the next data point. This process is repeated by updating the training set each time with the most recent data. This helps the model adapt to changes in the time series pattern, giving us a better idea of its predictive ability in real-time scenarios \cite{brownlee2017introduction}. The hyperparameters that yield the optimal average result across all validation windows are then selected for the ML prediction models.

A comparative analysis of various DL prediction algorithms is done. The comparison primarily focuses on evaluating prediction accuracy of the different algorithms. The prediction accuracy is assessed using various metrics, including the R2 Score, Mean Square Error (MSE), Root Mean Square Error (RMSE) and Mean Absolute Error (MAE) \cite{sone2020wireless,woodward2022time}. The R2 Score, also known as the coefficient of determination, is a statistical measure that quantifies the proportion of the variability in the observed time series that is explained by a predictive model. It measures the model’s goodness of fit by comparing the differences between the actual data and the model’s predictions. It provides insight into the model’s goodness of fit: a higher R2 score indicates that the model accounts for more of the variability in the data, while a lower score suggests that the model may not be a good fit for the data. It ranges from 0 to 1, representing no-fit to perfect fit, respectively. MSE measures the squared average of the differences between predicted and actual values. RMSE, is the square root of MSE. On the other hand, MAE calculates the absolute average of the residuals by summing the differences and dividing the result by the total number of samples. These metrics collectively provide comprehensive insights into the accuracy of the prediction models.

Moreover, the different prediction methods are compared in terms of Computational Time (CT). On the one hand, the CT in the phase of training (TCT) per second is computed. On the other hand, we determined the training runtime associated with online prediction (PCT) per second. By incorporating both training and test runtime in our analysis, we gain valuable insights into the computational cost of the different proposed prediction methodologies. The CT of the proposed methodologies is estimated using a computer with the following specifications: Intel(R) Core(TM) i5-11500 CPU @ 2.70GHz, 16.00 GB RAM, x64 based processor. 

Hyperparameter tuning is a crucial step in optimizing the prediction accuracy of DL algorithms. However, when dealing with time series NN models, exploring a large combination of hyperparameters can significantly slow down the training process \cite{brownlee2018deep}. Therefore, we focused on tuning the hyperparameters that have the most significant impact on prediction accuracy. Our approach involved experimenting with various combinations of hyperparameters to identify the optimal ones for specific performance metrics. Initially, we started with commonly used hyperparameters for time series forecasting as documented in \cite{sone2020wireless, brownlee2018deep, huang2017study}. In CNN model, we tested different combinations of the number of kernels \{16, 32, 64, 128, 256\} and the number of layers \{1, 2, 3\} to determine the best combination. Among these, a configuration with 16 kernels and 2 layers exhibited the highest accuracy across the experiments. Similarly, in the RNN model, specifically using LSTM, we explored combinations of the number of cells \{32, 50, 64, 100, 128, 256\} and the number of layers \{1, 2, 3\}. Notably, a configuration with 50 cells in each of the 2 layers demonstrated superior performance and was chosen as the optimal configuration. Finally, we optimized the hyperparameters for the CNN-RNN model by experimenting with different combinations of batch sizes \{16, 32\} and numbers of epochs \{50, 100\}. Among these combinations, the one that achieved the highest performance comprised a batch size of 16 and 100 epochs. The details of all the hyperparameters considered in the NN models are presented in Table \ref{tab:c3table1}

\begin{table}[t]
  \centering
  \scriptsize 
  \setlength{\tabcolsep} {4pt} 
  \caption{Hyperparameters of the CNN-RNN Algorithms (RNN refers to either SRNN, LSTM or GRU).}
  \label{tab:c3table1}
  \resizebox{\columnwidth}{!}{
      \begin{tabular}{|c|c|c|c|c|c|p{1cm}|}
        \hline
        \multirow{5}{*}{CNN} & Layers & Kernels & Kernel Size & Pool./Stride Size & AF  \\
        \cline{2-6}
        & Conv1D.1 & 16 & 2 & - & RelU\\
        \cline{2-6}
        & Pooling1 & - & - & Max/2 & - \\
        \cline{2-6}
        & Conv1D.2 & 16 & 2 & - & RelU \\
        \cline{2-6}
        & Pooling2 & - & - & Max/1 & - \\
        \hline
        \multirow{3}{*}{RNN} & Layers & Cells & Cell Type & Dropout Layer & AF  \\
        \cline{2-6} 
        & Layer1 & 50 & RNN & 0.3 & RelU\\
        \cline{2-6} 
        & Layer2 & 50 & RNN & 0.3 & RelU \\
        \hline
      \end{tabular}
  }
\end{table}

In our implementation, we considered hyperparameters similar to those described in the paper by \cite{zhou2021informer}. The Transformer architecture consists of 2 encoder layers and 1 decoder layer. The optimization process employs the ‘Adam’ optimizer, with an initial learning rate of $e^{-4}$, decaying by 0.5 after each epoch. For the transformer, we set the batch size to 16, the d\_model (dimension of the model) to 512, the number of attention head (\text{n\_head}) to 8, the dropout rate to 0.3 and the dimension of the feed-forward layer (\text{d\_ff}) to 2048. The training process is performed over approximately 10 epochs with appropriate early stopping mechanisms (patience = 3) to prevent overfitting and ensure optimal model performance.

\section{Obtained Results}
\label{secresults}
\subsection{Overview}
This section presents the obtained results for both the proposed temporal and spatio-temporal prediction methodologies. In order to illustrate the performance of the proposed methodologies, first, a simple scenario consisting of 6 APs located on the same floor (see Fig. \ref{fig:c3figure5}) is considered. Subsequently, the analysis is extended to the 100 APs located in seven different buildings on the University Campus. The prediction accuracy, the CT of the training phase (TCT) and the CT of the prediction phase (PCT) have been determined for all the proposed methodologies. Training Computational Time (TCT) consists of two distinct phases: (i) preprocessing including the computation of traffic load and failures measurements from raw data, conversion to the frequency of 30 minutes and backfilling, Augmented Dickey-Fuller test and (ii) training with walk-forward validation. 
Prediction Computational Time (PCT) encompasses only the duration of the prediction phase. After analysing the obtained results, it has been observed that all the considered methods provide the prediction with a small PCT which allows a prediction almost in real time. For this reason, the results provided in section \ref{c3sub5.2} and \ref{c3sub5.3}. According to this, section \ref{c3sub5.2} presents the obtained results when running only the temporal prediction. Then, section \ref{c3sub5.3} presents the results of the spatio-temporal predictions.

\subsection{Temporal Prediction}
\label{c3sub5.2}
The prediction results for the deployed 100 APs when running the temporal based prediction are presented in Table \ref{tab:c3table2} and Table \ref{tab:c3table3} for traffic load and transmission failures, respectively. In this case, the obtained prediction metrics correspond to the averaged values of each metric for all the APs. As shown in the tables, all the methods yield highly accurate predictions. LSTM stands out slightly with better performance compared to other algorithms, although GRU closely follows LSTM and exhibits slightly better performance in MAE. As depicted, a 89.73\% and 89.87\% R2 score is obtained for traffic load and failures prediction, respectively. LSTM provides the best prediction accuracy since it has an intrinsic ability to retain memory and capture sequential dependencies that allows it to model these complex relationships effectively. Note also that TCT devoted to training the model is relatively low (in the order of tenths or hundreds of seconds) for all the methods. 
As shown in the tables, the Transformer methodology exhibits a marginally lower prediction accuracy (e.g. R2 score) compared to the other methods and requires a much longer computational time. Note also that CNN has a relatively low TCT but provides the worse prediction accuracy because CNN is more adequate for space-based predictions and it is not so suitable for temporal-based prediction. In addition, the performance of all DL prediction models is compared with the ARIMA model. As shown, ARIMA’s predictions are slightly worse than those of DL algorithms, except for CNN. However, ARIMA takes more TCT to compute its predictions compared to all DL algorithms. This is mainly because ARIMA uses the technique of recursive sampling to improve its performance. Moreover, there is a noticeable difference in the ARIMA’s TCT for the case of traffic load and failures prediction. This difference is due to the fact that finding optimal order for each target AP measurements to achieve accurate predictions requires more TCT for traffic load than for failures. It is worth mentioning that between 40 and 50 seconds are spent on determining the best order using auto-arima function. 

According to PCT, Table \ref{tab:c3table2} and Table \ref{tab:c3table3} reveal that ARIMA exhibits the lowest PCT, while Transformer displays the highest PCT due to its more intricate architecture comparing to other methods. For NNs, the prediction time shown in the tables, excluding preprocessing, is very low, typically ranging between 0.20 to 0.30 seconds. However, the preprocessing time, which typically falls between 0.6 to 0.8 seconds, needs to be added to the prediction time. Consequently, the overall prediction time remains relatively low.

\begin{table}
  \centering
  \scriptsize 
  \setlength{\tabcolsep}{4pt} 
  \caption{Comparison among different Prediction Techniques for all 100 APs Traffic time series data.}
  \label{tab:c3table2}
  \resizebox{\columnwidth}{!}{
      \begin{tabular}{| c | c | c | c | c | c | c | p{1m}|}      
        \hline
         & R2 Score & MSE & RMSE & MAE & TCT(s)& PCT(s) \\
        \hline
         ARIMA & 0.8646 & 0.00036 & 0.01889 & 0.01641 & 427.072 & 0.00024 \\ 
        \hline
         SRNN & 0.8826 & 0.00020 & 0.01428 & 0.00830  & 69.414 & 0.245\\ 
        \hline
         \textbf{LSTM} & \textbf{0.8973} & \textbf{0.00019} & \textbf{0.01364} & 0.00813  & 110.862 & 0.300\\ 
        \hline
         GRU & 0.8968 & 0.00020 & 0.01421 & \textbf{0.00792} & 110.204 & 0.296 \\ 
        \hline
         CNN & 0.7821 & 0.00084 & 0.02896 & 0.01671 & 58.258 & 0.204\\ 
        \hline
         Trans. & 0.8704 & 0.00035 & 0.01881 & 0.01630 & 410.244 & 1.360\\ 
        \hline
      \end{tabular}
  }
\end{table}

\begin{table}
  \centering
  \scriptsize 
  \setlength{\tabcolsep}{4pt} 
  \caption{Comparison among different Prediction Techniques for all 100 APs Failures time series data.}
  \label{tab:c3table3}
  \resizebox{\columnwidth}{!}{
      \begin{tabular}{| c | c | c | c | c | c | c |p{1cm}|}      
        \hline
         & R2 Score & MSE & RMSE & MAE & TCT(s) & PCT(s) \\
        \hline
         ARIMA  &   0.8745  &   0.00386  &   0.06216  &   0.04921  &   325.374 & 0.00025\\ 
        \hline
         SRNN  &   0.8830  &   0.00336  &   0.05795  &   0.04221  & 70.295 & 0.248\\ 
        \hline
         \textbf{LSTM}  &   \textbf{0.8987}  &   \textbf{0.00254}  &   \textbf{0.05042}  &   0.03854  &   112.134 & 0.303 \\ 
        \hline
         GRU  &   0.8942  & 0.00257  &   0.05068  & \textbf{0.03842} & 110.358 & 0.300 \\ 
        \hline
         CNN  &   0.7397  &   0.00686  &   0.08281  &   0.05958  &  59.318 & 0.207\\ 
        \hline
         Trans.  &   0.8872  &  0.00313  & 0.05592  & 0.03885  &  411.755  & 1.462 \\ 
        \hline
       \end{tabular}
   }
\end{table}

\begin{figure*}[t]
  \centering
  \includegraphics[width=1\linewidth]{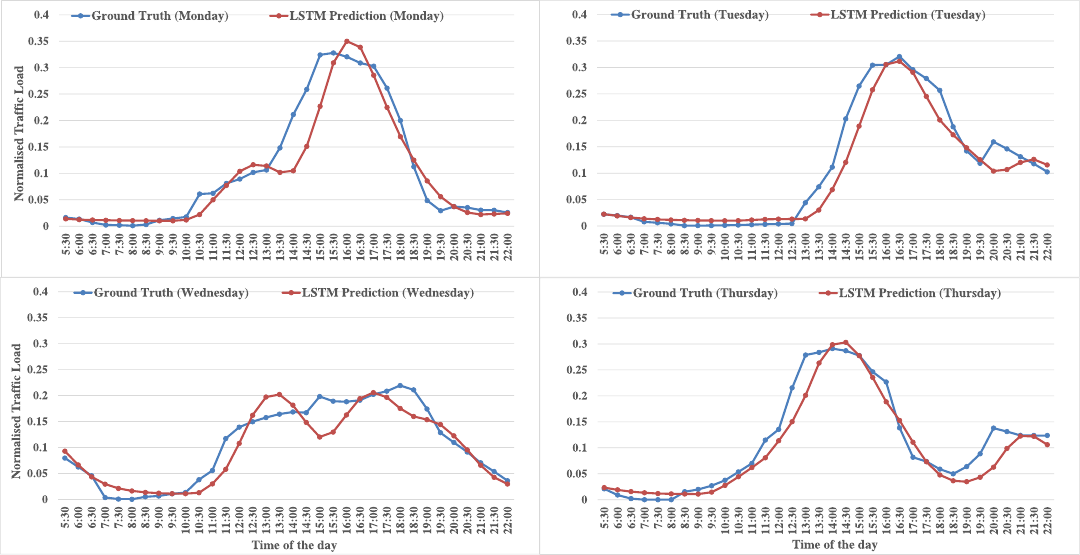}  
  \caption{Comparing the Real and Prediction traffic load of target AP(XSFA4PS205) using LSTM in four days of the last week.}
  \label{fig:c3figure6}
\end{figure*}

Fig. \ref{fig:c3figure6} illustrates a comparison between real values and obtained predictions using LSTM during four days of the last week in Autumn for single AP (XSFA4PS205). As shown, most of the prediction measurements are quite close to real data.
\subsection{Spatio-temporal Prediction}
\label{c3sub5.3}
In order to illustrate the performance of the spatio-temporal prediction methodology in a relatively simple scenario, let consider the scenario of 6 APs described in Fig. \ref{fig:c3figure5}. These 6 APs have automatically been included in the neighbouring list of each other based on the threshold of -80dBm specified by Cisco Prime Infrastructure. Then, the Pearson correlation \cite{wang2017spatiotemporal,sone2020wireless} among the 6 APs are presented in Table \ref{tab:c3table4} and Table \ref{tab:c3table5}. Considering the prediction of traffic and failures for XSFA4PS205, the methodology determines the highly correlated neighbour APs as those having a correlation $C_{i,j}> 0.50$. This threshold has been empirically set to maximize the obtained R2 score. In the example shown in Table \ref{tab:c3table4}, AP XSFA4PS205 and its 4 highly correlated neighbour APs (i.e. XSFA4PS201, XSFA4PS203, XSFA4PS204 and XSFA4PS206) are used for spatio-temporal prediction of traffic data. However, for the case of spatio-temporal prediction of failures in XSFA4PS205 all 5 neighbouring APs are considered, as shown in Table \ref{tab:c3table5}.

\begin{table}[t]
  \centering
  \scriptsize 
  \setlength{\tabcolsep}{4pt} 
  \caption{Spatial Correlation for Traffic data (In order to simplification, the prefix "XSFA4PS" at the beginning of the AP name has been omitted).}
  \label{tab:c3table4}
  \resizebox{\columnwidth}{!}{
  \begin{tabular}{| c | c | c | c | c | c | c |p{1cm}|}        
    \hline
    \textbf{Traffic} & 201 & 202 & 203 & 204 & 205 & 206 \\
    \hline
    201  &   1  &   0.4950  &   0.5622  &   0.6824  &   0.5101  &   0.4907  \\ 
    \hline
    202  &   0.4950  &   1  &   0.4185  &   0.5042  &   0.4172  &   0.4356  \\ 
    \hline
    203  &   0.5622  &   0.4185  &   1  &   0.5985  &   0.5489  &   0.5049  \\ 
    \hline
    204  &   0.6824  &   0.5042  &   0.5985  &   1  &   0.5384  &   0.4729  \\ 
    \hline
    205  &   0.5101  &   0.4172  &   0.5489  &   0.5384  &   1  &   0.5130  \\ 
    \hline
    206  &   0.4907  &   0.4356  &   0.5049  &   0.4729  &   0.5130  &   1  \\ 
    \hline
  \end{tabular}
  }
\end{table}

\begin{table}[t]
  \centering
  \scriptsize 
  \setlength{\tabcolsep}{4pt} 
  \caption{Spatial Correlation for Failures data (In order to simplification, the prefix "XSFA4PS" at the beginning of the AP name has been omitted).}
  \label{tab:c3table5}
  \resizebox{\columnwidth}{!}{
  \begin{tabular}{| c | c | c | c | c | c | c |p{1cm}|}       
    \hline
    \textbf{Failures} & 201 & 202 & 203 & 204 & 205 & 206 \\
    \hline
    201  &   1  &   0.7122  &   0.8494  &   0.8612  &   0.7659  &   0.5229  \\ 
    \hline
    202  &   0.7122  &   1  &   0.7123  &   0.7239  &   0.7038  &   0.4849  \\ 
    \hline
    203  &   0.8494  &   0.7123  &   1  &   0.8221  &   0.7831  &   0.6046  \\ 
    \hline
    204  &   0.8612  &   0.7239  &   0.8221  &   1  &   0.7985  &   0.5508  \\ 
    \hline
    205  &   0.7659  &   0.7038  &   0.7831  &   0.7985  &   1  &   0.6041  \\ 
    \hline
    206  &   0.5229  &   0.4849  &   0.6046  &   0.5508  &   0.6041  &   1  \\ 
    \hline
  
  \end{tabular}
  }
\end{table}

Based on this, the results obtained with the spatio-temporal prediction methodology are presented in Table \ref{tab:c3table6} and Table \ref{tab:c3table7} and are also compared with the results of temporal prediction of this AP. In terms of temporal prediction, LSTM demonstrates superior performance compared to the other methods, with the exception of MAE for traffic prediction, where GRU and LSTM exhibit similar performance, albeit with a slightly better performance in GRU. For benchmarking purposes, in spatio-temporal context, the first three methods presented in the Tables (i.e. LSTM, GRU and CNN) correspond to single DL algorithms while the last three methods, (CNN-SRNN, CNN-LSTM and CNN-GRU) correspond to hybrid CNN-RNN prediction algorithms. As shown in Tables \ref{tab:c3table6} and \ref{tab:c3table7}, the results obtained by the spatio-temporal LSTM, CNN and GRU methodologies provide higher prediction accuracy at the expenses of an increase in the TCT with respect to the methodologies based on only temporal prediction. In any case, this increase in TCT is not relevant since the training is done offline. The online computational time, PCT for both temporal and spatio-temporal is approximately similar. In Table \ref{tab:c3table6} and Table \ref{tab:c3table7}, the Transformer’s accuracy closely aligns with that of GRU and LSTM, respectively. Focusing on the hybrid CNN-RNN models based on the architecture shown in Fig. \ref{fig:c3figure3}, it is worth noting that they provide a higher prediction performance with a lower TCT when compared to the single DL methods (see Tables \ref{tab:c3table6} and \ref{tab:c3table7}). The reason is that the inclusion of the first CNN process in the hybrid methodology, in order to find the spatial relationships among the different APs, reduces the number of features after 1D-Conv and pooling layer, thus reducing the computational cost of the latter RNN process. As depicted in the tables, the hybrid CNN-LSTM model achieves the highest prediction accuracy for both traffic usage and failures of a target AP, except for MAE in traffic load, where CNN-GRU slightly outperforms CNN-LSTM (see Table \ref{tab:c3table6}). Note also that CNN-SRNN provides a relatively high prediction accuracy (e.g. in terms of R2 score) with a considerably lower TCT than CNN-LSTM. The PCT for both single and hybrid NN models exhibits minimal difference or is closely matched. The PCT of the Transformer model is higher in spatio-temporal prediction compared to other models. However, when compared to temporal prediction, it is relatively similar or close.

\begin{table}[t]
  \centering
  \scriptsize 
  \setlength{\tabcolsep}{4pt} 
  \caption{Comparison of different Traffic Prediction Methods for AP XSFA4PS205.}
  \label{tab:c3table6}
  \resizebox{\columnwidth}{!}{
      \begin{tabular}{|c|c|c|c|c|c|c|c|c|p{1cm}|}
        \hline
        & \multicolumn{2}{|c|}{DL Methods} & R2 Score & MSE & RMSE & MAE & TCT(s) & PCT(s) \\
        \hline
        \multirow{5}{*}{\rotatebox[origin=c]{90}{Temporal}{\begin{tabular}{@{}c@{}}{\rotatebox[origin=c]{90}{Prediction}}\end{tabular}}} & \multirow{5}{*}{\rotatebox[origin=c]{90}{Single}} & SRNN & 0.8737 & 0.00121 & 0.03479 & 0.02479 & 71.3415 & 0.267  \\
        \cline{3-9}
        & & \textbf{LSTM} & \textbf{0.8957} & \textbf{0.00099} & \textbf{0.03162} & 0.02207 & 106.557 & 0.305\\
        \cline{3-9}
        & & GRU & 0.8908 & 0.00105 & 0.03235 & \textbf{0.02184} & 107.354 & 0.305 \\
        \cline{3-9}
        & & CNN & 0.8754 & 0.00119 & 0.03456 & 0.02608 & 57.164 & 0.207 \\
        \cline{3-9}
        & & Transformer & 0.8670 & 0.00128 & 0.03574 & 0.02495 & 406.238 & 1.521 \\
        \hline
        \multirow{7}{*}{\rotatebox[origin=c]{90}{Spatio-temporal}{\begin{tabular}{@{}c@{}}{\rotatebox[origin=c]{90}{Prediction}}\end{tabular}}} & \multirow{4}{*}{\rotatebox[origin=c]{90}{Single}} & LSTM & 0.9241 & 0.00073 & 0.02698 & 0.01966 & 248.245 & 0.302  \\
        \cline{3-9}
        & & GRU & 0.9197 & 0.00077 & 0.02775 & 0.01971 & 238.142 & 0.286 \\
        \cline{3-9}
        & & CNN & 0.8854 & 0.00100 & 0.03164 & 0.02605 & 98.455 & 0.187 \\
        \cline{3-9}
        & & Transformer & 0.9174 & 0.00082 & 0.02868 & 0.02073 & 523.15 & 1.891 \\
        \cline{2-9}
        & \multirow{3}{*}{\rotatebox[origin=c]{90}{Hybrid}} & CNN-SRNN & 0.9350 & 0.00062 & 0.02496 & 0.01860 & 152.778 & 0.234 \\
        \cline{3-9}
        & & \textbf{CNN-LSTM} & \textbf{0.9521} & \textbf{0.00046} & \textbf{0.02143} & 0.01612 & 233.547 & 0.271 \\
        \cline{3-9}
        & & CNN-GRU & 0.9465 & 0.00051 & 0.02262 & \textbf{0.01599} & 223.049 & 0.264\\
        \hline
      \end{tabular}
  }
\end{table}

\begin{table}[t]
  \centering
  \scriptsize 
  \setlength{\tabcolsep}{4pt} 
  \caption{Comparison of different Failures Prediction Methods for AP XSFA4PS205.}
  \label{tab:c3table7}
  \resizebox{\columnwidth}{!}{
      \begin{tabular}{|c|c|c|c|c|c|c|c|c|p{1cm}|}
        \hline
        & \multicolumn{2}{|c|}{DL Methods} & R2 Score & MSE & RMSE & MAE & TCT(s) & PCT(s)\\
        \hline
        \multirow{5}{*}{\rotatebox[origin=c]{90}{Temporal}{\begin{tabular}{@{}c@{}}{\rotatebox[origin=c]{90}{Prediction}}\end{tabular}}} & \multirow{5}{*}{\rotatebox[origin=c]{90}{Single}} & SRNN & 0.8257 & 0.00939 & 0.09692 & 0.07217 & 72.159 & 0.256\\
        \cline{3-9}
        & & \textbf{LSTM} & \textbf{0.9073} & \textbf{0.00498} & \textbf{0.07060} & \textbf{0.05607} & 108.337 & 0.307\\
        \cline{3-9}
        & & GRU & 0.8896 & 0.00594 & 0.07708 & 0.05750 & 105.474 & 0.298 \\
        \cline{3-9}
        & & CNN & 0.8700 & 0.00699 & 0.08361 & 0.06699 & 57.170 & 0.254\\
        \cline{3-9}
        & & Transformer & 0.9044 & 0.00506 & 0.07116 & 0.05614 & 406.684 & 1.421\\
        \hline
        \multirow{7}{*}{\rotatebox[origin=c]{90}{Spatio-temporal}{\begin{tabular}{@{}c@{}}{\rotatebox[origin=c]{90}{Prediction}}\end{tabular}}} & \multirow{4}{*}{\rotatebox[origin=c]{90}{Single}} & LSTM & 0.9286 & 0.00382 & 0.06177 & 0.04748 & 254.754 & 0.261\\
        \cline{3-9}
        & & GRU & 0.9118 & 0.00471 & 0.06865 & 0.05178 & 247.456 & 0.248\\
        \cline{3-9}
        & & CNN & 0.9041 & 0.00513 & 0.07157 & 0.05555 & 116.354 & 0.203\\
        \cline{3-9}
        & & Transformer & 0.9265 & 0.00404 & 0.06354 & 0.05004 & 534.799 & 1.721\\
        \cline{2-9}
        & \multirow{3}{*}{\rotatebox[origin=c]{90}{Hybrid}} & CNN-SRNN & 0.9551 & 0.00239 & 0.04895 & 0.03864 & 157.246 & 0.252\\
        \cline{3-9}
        & & \textbf{CNN-LSTM} & \textbf{0.9599} & \textbf{0.00214} & \textbf{0.04631} & \textbf{0.03698} & 238.472 & 0.302\\
        \cline{3-9}
        & & CNN-GRU & 0.9523 & 0.00255 & 0.05048 & 0.03919 & 237.576 & 0.294\\
        \hline
      \end{tabular}
  }
\end{table}

It's important to note that having a higher number of correlated neighbours may not necessarily lead to higher accuracy in predictions. The quality of these correlations, meaning how strong they are, is also crucial. To illustrate this point, we consider a quantitative example involving a target AP of XSFA4PS205 in failure measurements. As Table \ref{tab:c3table5} shows, by adjusting the correlation threshold from 0.50 to 0.77, we will have 2 highly correlated APs for the target AP. We then evaluate the performance of a CNN-LSTM model on failure metrics across this new threshold. Table \ref{tab:c3threshold} observes a slight improvement in results as we increase the threshold and consequently reduce the number of correlated APs. However, a too high correlation threshold may lead to situations in which an AP may not have any neighbours with such a high correlation. This may cause that relevant information of neighbouring APs is not considered for prediction leading to worse prediction results. In order to deal with this trade-off, we have selected a correlation threshold of 0.5.

\begin{table}[t]
  \centering
  \scriptsize 
  \setlength{\tabcolsep}{12pt} 
  \caption{Comparison between two threshold ($C_i,j$)s for AP XSFA4PS205.}
  \label{tab:c3threshold}
  \resizebox{\columnwidth}{!}{
  \begin{tabular}{| c | c | c | c | c |p{1cm}|}       
    \hline
    \textbf{$C_i,j$} & R2 Score & MSE & RMSE & MAE \\
    \hline
    0.5  &   0.9599  &   0.00214  &   0.04631  &   0.03698  \\ 
    \hline
    0.77  &   0.9684  &   0.00167  &   0.04083  &   0.03252  \\ 
    \hline
     
  \end{tabular}
  }
\end{table}

The proposed spatio-temporal prediction methodologies have also been evaluated for all the 100 APs located in seven different buildings. After the analysing the traffic correlation among the different APs it is observed that 42APs have one or more neighbouring APs with a correlation $C_{i,j}$ higher than the established threshold (i.e. $C_{i,j}= 0.50 $). For the case of the failures’ correlations among APs, 55APs with one or more highly correlated neighbours are identified. In order to compare the different spatio-temporal prediction DL methods, Table \ref{tab:c3table8} and Table \ref{tab:c3table9} present the obtained results when running the spatio-temporal prediction methodology only for the APs with one or more highly correlated neighbour (i.e. 42 APs for traffic predictions and 55APs for failure predictions). The obtained prediction metrics correspond to the averaged values of each metric for all the APs. Note that, according to Table \ref{tab:c3table8} and Table \ref{tab:c3table9}, the hybrid DL methods outperform the prediction accuracy metrics with respect to single DL models. The average TCT for spatio-temporal prediction across 42 APs for traffic load and 55 APs for failures is relatively similar to that of AP XSFA4PS205. The PCT for NN methods ranges between 0.20 and 0.32 seconds, whereas Transformers exhibit a higher PCT compared to others.

\begin{table}[t]
  \centering
  \scriptsize 
  \setlength{\tabcolsep}{4pt} 
  \caption{Comparison of different Traffic Prediction Methods for the 42 APs with one or more highly correlated neighbours.}
  \label{tab:c3table8}
  \resizebox{\columnwidth}{!}{
      \begin{tabular}{|c|c|c|c|c|c|c|p{1cm}|}
        \hline
        \multicolumn{2}{|c|}{DL Methods} & R2 Score & MSE & RMSE & MAE & TCT(s) & PCT(s) \\
        \hline
        \multirow{4}{*}{\rotatebox[origin=c]{90}{Single}} & LSTM & 0.9250 & 0.00061 & 0.02464 & 0.01518 & 247.488 & 0.301  \\
        \cline{2-8}
        & GRU & 0.9202 & 0.00050 & 0.02240 & 0.01482 & 241.516 & 0.313 \\
        \cline{2-8}
        & CNN & 0.8471 & 0.00114 & 0.03378 & 0.02186 & 100.225 & 0.206\\
        \cline{2-8}
        & Transformer & 0.8930 & 0.00080 & 0.02834 & 0.01962 & 435.09 & 1.457\\
        \hline
        \multirow{3}{*}{\rotatebox[origin=c]{90}{Hybrid}} & CNN-SRNN & 0.9313 & 0.00057 & 0.02392 & 0.01483 & 153.232 & 0.233 \\
        \cline{2-8}
        & CNN-LSTM & \textbf{0.9371} & \textbf{0.00047} & \textbf{0.02161} & \textbf{0.01364} & 235.086 & 0.273 \\
        \cline{2-8}
        & CNN-GRU & 0.9345 & 0.00050 & 0.02243 & 0.01422 & 225.047 & 0.271 \\
        \hline
      \end{tabular}
  }
\end{table}

\begin{table}[t]
  \centering
  \scriptsize 
  \setlength{\tabcolsep}{4pt} 
  \caption{Comparison of different Failures Prediction Methods for the 55 APs with one or more highly correlated neighbours.}
  \label{tab:c3table9}
  \resizebox{\columnwidth}{!}{
      \begin{tabular}{|c|c|c|c|c|c|c|p{1cm}|}
        \hline
        \multicolumn{2}{|c|}{DL Methods} & R2 Score & MSE & RMSE & MAE & TCT(s) & PCT(s)\\
        \hline
        \multirow{4}{*}{\rotatebox[origin=c]{90}{Single}} & LSTM & 0.9065 & 0.00278 & 0.05270 & 0.04141 & 248.958 & 0.259  \\
        \cline{2-8}
        & GRU & 0.9058 & 0.00268 & 0.05178 & 0.04151 & 243.783 & 0.258 \\
        \cline{2-8}
        & CNN & 0.8046 & 0.00574 & 0.07579 & 0.05967 & 114.481 & 0.207 \\
        \cline{2-8}
        & Transformer & 0.9050 & 0.00270 & 0.05200 & 0.04096 & 486.110 & 1.768\\
        \hline
        \multirow{3}{*}{\rotatebox[origin=c]{90}{Hybrid}} & CNN-SRNN & 0.9081 & 0.00268 & 0.05178 & 0.04073 & 153.893 & 0.240 \\
        \cline{2-8}
        & CNN-LSTM & \textbf{0.9241} & \textbf{0.00223} & \textbf{0.04720} & \textbf{0.03727} & 234.192 & 0.271 \\
        \cline{2-8}
        & CNN-GRU & 0.9229 & 0.00225 & 0.04745 & 0.03733 & 232.667 & 0.251 \\
        \hline
      \end{tabular}
  }
\end{table}

\begin{table*}[t]
  \centering
  \scriptsize 
  \setlength{\tabcolsep}{3pt} 
  \caption{Comparison of different Proposed Temporal and Spatio-temporal Prediction Methodologies for Traffic Load and Failures based on LSTM.}
  \label{tab:c3table10}
      \begin{tabular}{|c|c|c|c|c|c|c|c|c|c|c|p{1cm}|}
        \hline
        \multicolumn{2}{|c|}{} & \multicolumn{3}{|c|}{Temporal Prediction} & \multicolumn{6}{|c|}{Spatio-temporal Prediction}\\
        \cline{3-11}
        \multicolumn{2}{|c|}{} & \multicolumn{3}{|c|}{LSTM} & \multicolumn{3}{|c|}{Single LSTM} & \multicolumn{3}{|c|}{Hybrid CNN-LSTM} \\
        \hline
        Considered APs in the analysis & Metric Prediction & R2 Score & TCT(s) & PCT(s) & R2 Score & TCT(s) & PCT(s) & R2 Score & TCT(s) & PCT(s)\\
        \hline
        42 APs & Traffic Load & 0.9175 & 111.700 & 0.298 & 0.9250 & 247.488 & 0.301 & 0.9371 & 235.086 & 0.273 \\
        \hline
        55 APs & Tx. Failures & 0.9043 & 113.972 & 0.304 & 0.9065 & 248.958 & 0.259 & 0.9241 & 234.192 & 0.271 \\
        \hline
      \end{tabular}
\end{table*}

\begin{table*}[t]
  \centering
  \scriptsize 
  \setlength{\tabcolsep}{3pt} 
  \caption{Performance of the Combined Temporal and Spatio-temporal Prediction Methodology for Traffic Load and Failures based on LSTM.}
  \label{tab:c3table11}
      \begin{tabular}{|c|c|c|c|c|c|c|c|c|c|c|p{1cm}|}
        \hline
        \multicolumn{2}{|c|}{} & \multicolumn{3}{|c|}{Temporal Prediction} & \multicolumn{6}{|c|}{Combined Temoporal and Spatio-temporal Prediction}\\
        \cline{3-11}
        \multicolumn{2}{|c|}{} & \multicolumn{3}{|c|}{LSTM} & \multicolumn{3}{|c|}{Single LSTM} & \multicolumn{3}{|c|}{Hybrid CNN-LSTM} \\
        \hline
        Considered APs in the analysis & Metric Prediction & R2 Score & TCT(s) & PCT(s) & R2 Score & TCT(s) & PCT(s) & R2 Score & TCT(s) & PCT(s) \\
        \hline
        100 APs & Traffic Load & 0.8973 & 110.862 & 0.300 & 0.8989 & 171.265 & 0.299 & 0.9041 & 165.782 & 0.284 \\
        \hline
        100 APs & Tx. Failures & 0.8987 & 112.134 & 0.303 & 0.9000 & 191.341 & 0.271 & 0.9101 & 182.795 & 0.287 \\
        \hline
      \end{tabular}
\end{table*}
Finally, Table \ref{tab:c3table10} and Table \ref{tab:c3table11} summarize the performance of the different prediction methodos when considering the LSTM algorithm. In particular Table \ref{tab:c3table10} compares the different proposed temporal and spatio-temporal predictions for the 42 APs that have one or more highly traffic correlated neighbours and the 55 APs that have one or more highly failure-correlated neighbours. As shown, the proposed hybrid spatio-temporal prediction methods, specifically CNN-LSTM, provide slightly higher prediction accuracy (i.e. R2 Score) at the expense of a increase in the average TCT with respect to the only temporal prediction for both traffic load and transmission failures. The PCT values for all methods range from 0.259 to 0.304 seconds, indicating relatively low computational overhead. 

Table \ref{tab:c3table11} illustrates the performance of a combined methodology that decides, for each AP, whether to run the temporal prediction (with LSTM algorithm) or the spatio-temporal prediction (with single LSTM or with hybrid CNN-LSTM) depending on the obtained results of the neighbour correlation analysis. For comparison purposes, the only temporal-based predictions with the LSTM algorithm are also presented in Table \ref{tab:c3table11}. This comparison is done for all the 100 APs. As illustrated, the combined temporal LSTM and spatio-temporal CNN-LSTM have slightly better prediction accuracy than the only temporal LSTM at the cost of a higher TCT. However, the PCT values for both the temporal prediction, and combined temporal and spatio-temporal prediction models remain very similar.

\section{Implementation Aspects}
\label{secImpleAspects}
The imperative need for improved network management practices in Wi-Fi environments, underscored by the rapid development and increased usage of Wi-Fi technologies, has motivated the development of a methodology for predicting future network metrics. By accurately forecasting the network metrics such as traffic load and transmission failures, this methodology aims to optimize resource allocation, proactively address maintenance issues, and plan for future capacity needs. This proactive approach not only enhances network reliability and performance but also improves the overall user experience by minimizing downtime and congestion. Additionally, the potential for cost optimization through reduced reactive maintenance and informed resource investments further highlights the significance of this research in addressing real-world challenges in network management.

Implementing our methodology in a real network involves defining and addressing Key Performance Indicators (KPIs) that are relevant to the specific network environment and operational requirements. KPIs such as TCT, PCT, prediction accuracy are essential metrics to consider when evaluating the performance of prediction methods in a real network setting.  In general, various methods were assessed for temporal and spatio-temporal prediction in a Wi-Fi network. The TCT for training the models was relatively low across all methods, facilitating fast offline training and retraining processes. This ensures that models can be updated frequently, if needed, while maintaining low computational costs. Furthermore, the PCT for both temporal and spatio-temporal prediction was found to be minimal, allowing for almost real-time predictions.

For the scalability involving a greater number of APs (i.e. addition of new APs), it is necessary to collect data for each new AP and its neighbours, pre-process (fill missing data, etc.), calculate correlations and execute trained model. Notably, the training of the model, which constitutes a significant portion of the computational burden, does not need to be repeated. 
Furthermore, in our scenario, adapting to new circumstances may entail either fine-tuning the model's trainable parameters or completely retraining it. For example, in the considered scenario of a University Campus, when there are changes in classroom setups or new academic programs introduced at the university, adjusting the predictive model becomes necessary. This adjustment involves fine-tuning the model's parameters or retraining from scratch to better fit the new circumstances. In our data after splitting the training dataset to training and validation (70-30) by using walk-forward validation, the model considers about one month for validation. 
The trained model is valid during all the time period of our available dataset, but a retraining from scratch might be necessary in the future. In general, an evaluation of the frequency of retraining requires more available data, which is outlined as part of future work.

\section{Conclusion}
\label{secconclusion}
Prediction of future values of network metrics (such as future traffic values at the APs, future obtained performance metrics, etc.) can be useful to make proactive decisions over the network by means of e.g. load balancing, proactive resource allocation, congestion control, etc., that can improve the network performance. This paper has proposed a new methodology for the prediction of future values of some specific parameters according to historical measurements and by leveraging space correlations among neighbouring APs. The proposed methodology begins with a spatial correlation analysis of the collected measurements among a set of APs in a region. According to the result of this correlation analysis, it either performs a temporal prediction based on historical measurements of the target AP or a spatio-temporal prediction that incorporates historical measurements from the target AP and its highly correlated neighbouring APs. The proposed prediction methodology relies on Neural Networks and operates in two steps. First, the collected historical data is used to train the model in an offline process. Then, newly collected data is input into the trained model during the prediction step. 

Different DL methods have been analysed, including SimpleRNN, CNN, GRU, LSTM and Transformer for both temporal and spatio-temporal prediction tasks. Moreover, hybrid DL algorithms have been proposed, consisting of an initial CNN process to extract spatial correlations and a subsequent RNN step to exploit temporal correlations. The proposed methodology has been tested using real data obtained from a Wi-Fi network deployed on a University Campus to predict future AP traffic and transmission failures. The proposed methodology with the hybrid DL methods obtains high prediction accuracy (i.e. average R2 score up to 93.7\% and 92.4\% for traffic load and failures, respectively) with a relatively small TCT (i.e. in the order of hundreds of seconds) and a reduced PCT (lower than 2 seconds). In particular, exploiting the spatial correlation among neighbouring APs results in higher prediction accuracy for spatio-temporal predictions with respect to the only temporal predictions, albeit with a slight increase in the TCT. However, the PCT values for both the temporal and spatio-temporal prediction NN models remain very similar. Finally, the performance evaluation of the combined temporal and spatio-temporal prediction methodology illustrates slight improvements in prediction accuracy when exploiting the spatial domain for the APs with one or several highly correlated AP neighbours. 

The paper has also discussed the implementation aspects of our methodology in a real network by (i) comparing different ML techniques in terms of KPIs relevant to real network operation and management, such as accuracy, TCT, and PCT, to evaluate the benefits and drawbacks of each technique, and (ii) briefly discussing the scalability of a greater number of APs and a necessity of model retraining.

\section*{Acknowledgement}
This paper has been funded by the Spanish Ministry of Science and Innovation MCIN/AEI/10.13039/501100011033 under ARTIST project (ref. PID2020- 115104RB-I00).

 \bibliographystyle{elsarticle-num} 
 \bibliography{cas-refs}

\begin{thebibliography}{10}
\expandafter\ifx\csname url\endcsname\relax
  \def\url#1{\texttt{#1}}\fi
\expandafter\ifx\csname urlprefix\endcsname\relax\def\urlprefix{URL }\fi
\expandafter\ifx\csname href\endcsname\relax
  \def\href#1#2{#2} \def\path#1{#1}\fi

\bibitem{cisco2020cisco}
U.~Cisco, {Cisco annual internet report (2018--2023) white paper}, Cisco: San Jose, CA, USA 10~(1) (2020) 1--35.

\bibitem{gacanin2017wi}
H.~Gacanin, A.~Ligata, {Wi-Fi self-organizing networks: challenges and use cases}, IEEE Communications Magazine 55~(7) (2017) 158--164.

\bibitem{kibria2018big}
M.~G. Kibria, K.~Nguyen, G.~P. Villardi, O.~Zhao, K.~Ishizu, F.~Kojima, {Big data analytics, machine learning, and artificial intelligence in next-generation wireless networks}, IEEE access 6 (2018) 32328--32338.

\bibitem{zhang2019new}
K.~Zhang, G.~Chuai, W.~Gao, X.~Liu, S.~Maimaiti, Z.~Si, {A new method for traffic forecasting in urban wireless communication network}, EURASIP Journal on Wireless Communications and Networking 2019 (2019) 1--12.

\bibitem{wang2017spatiotemporal}
J.~Wang, J.~Tang, Z.~Xu, Y.~Wang, G.~Xue, X.~Zhang, D.~Yang, {Spatiotemporal modeling and prediction in cellular networks: A big data enabled deep learning approach}, in: IEEE INFOCOM 2017-IEEE conference on computer communications, IEEE, 2017, pp. 1--9.

\bibitem{thapaliya2018predicting}
A.~Thapaliya, J.~Schnebly, S.~Sengupta, {Predicting congestion level in wireless networks using an integrated approach of supervised and unsupervised learning}, in: 2018 9th IEEE Annual Ubiquitous Computing, Electronics \& Mobile Communication Conference (UEMCON), IEEE, 2018, pp. 977--982.

\bibitem{khan2020real}
M.~A. Khan, R.~Hamila, N.~A. Al-Emadi, S.~Kiranyaz, M.~Gabbouj, {Real-time throughput prediction for cognitive Wi-Fi networks}, Journal of Network and Computer Applications 150 (2020) 102499.

\bibitem{chen2021flag}
W.~Chen, F.~Lyu, F.~Wu, P.~Yang, J.~Ren, {Flag: Flexible, accurate, and long-time user load prediction in large-scale WiFi system using deep RNN}, IEEE Internet of Things Journal 8~(22) (2021) 16510--16521.

\bibitem{sone2020wireless}
S.~P. Sone, J.~J. Lehtom{\"a}ki, Z.~Khan, {Wireless traffic usage forecasting using real enterprise network data: Analysis and methods}, IEEE Open Journal of the Communications Society 1 (2020) 777--797.

\bibitem{sone2021forecasting}
S.~P. Sone, J.~Lehtom{\"a}ki, Z.~Khan, K.~Umebayashi, {Forecasting wireless network traffic and channel utilization using real network/physical layer data}, in: 2021 Joint European Conference on Networks and Communications \& 6G Summit (EuCNC/6G Summit), IEEE, 2021, pp. 31--36.

\bibitem{aceto2019mobile}
G.~Aceto, D.~Ciuonzo, A.~Montieri, A.~Pescap{\'e}, {Mobile encrypted traffic classification using deep learning: Experimental evaluation, lessons learned, and challenges}, IEEE Transactions on Network and Service Management 16~(2) (2019) 445--458.

\bibitem{aceto2019mimetic}
G.~Aceto, D.~Ciuonzo, A.~Montieri, A.~Pescap{\`e}, {MIMETIC: Mobile encrypted traffic classification using multimodal deep learning}, Computer networks 165 (2019) 106944.

\bibitem{deng2021short}
T.~Deng, M.~Wan, K.~Shi, L.~Zhu, X.~Wang, X.~Jiang, {Short term prediction of wireless traffic based on tensor decomposition and recurrent neural network}, SN Applied Sciences 3~(9) (2021) 779.

\bibitem{xu2017high}
Y.~Xu, W.~Xu, F.~Yin, J.~Lin, S.~Cui, {High-accuracy wireless traffic prediction: A GP-based machine learning approach}, in: GLOBECOM 2017-2017 IEEE Global Communications Conference, IEEE, 2017, pp. 1--6.

\bibitem{nie2017network}
L.~Nie, D.~Jiang, S.~Yu, H.~Song, {Network traffic prediction based on deep belief network in wireless mesh backbone networks}, in: 2017 IEEE Wireless Communications and Networking Conference (WCNC), IEEE, 2017, pp. 1--5.

\bibitem{li2020prophet}
Y.~Li, Z.~Ma, Z.~Pan, N.~Liu, X.~You, {Prophet model and Gaussian process regression based user traffic prediction in wireless networks}, Science China Information Sciences 63 (2020) 1--8.

\bibitem{wei2012intrusion}
M.~Wei, K.~Kim, {Intrusion detection scheme using traffic prediction for wireless industrial networks}, Journal of Communications and Networks 14~(3) (2012) 310--318.

\bibitem{zhang2018citywide}
C.~Zhang, H.~Zhang, D.~Yuan, M.~Zhang, {Citywide cellular traffic prediction based on densely connected convolutional neural networks}, IEEE Communications Letters 22~(8) (2018) 1656--1659.

\bibitem{huang2017study}
C.-W. Huang, C.-T. Chiang, Q.~Li, {A study of deep learning networks on mobile traffic forecasting}, in: 2017 IEEE 28th annual international symposium on personal, indoor, and mobile radio communications (PIMRC), IEEE, 2017, pp. 1--6.

\bibitem{li2020deep}
M.~Li, Y.~Wang, Z.~Wang, H.~Zheng, {A deep learning method based on an attention mechanism for wireless network traffic prediction}, Ad Hoc Networks 107 (2020) 102258.

\bibitem{wen2020assisting}
J.~Wen, M.~Sheng, J.~Li, K.~Huang, {Assisting intelligent wireless networks with traffic prediction: Exploring and exploiting predictive causality in wireless traffic}, IEEE Communications Magazine 58~(6) (2020) 26--31.

\bibitem{li2022lightweight}
G.~Li, S.~Zhong, X.~Deng, L.~Xiang, S.-H.~G. Chan, R.~Li, Y.~Liu, M.~Zhang, C.-C. Hung, W.-C. Peng, {A Lightweight and Accurate Spatial-Temporal Transformer for Traffic Forecasting}, IEEE Transactions on Knowledge and Data Engineering (2022).

\bibitem{hu2022citywide}
Y.~Hu, Y.~Zhou, J.~Song, L.~Xu, X.~Zhou, {Citywide Mobile Traffic Forecasting Using Spatial-Temporal Downsampling Transformer Neural Networks}, IEEE Transactions on Network and Service Management 20~(1) (2022) 152--165.

\bibitem{gu2023spatial}
B.~Gu, J.~Zhan, S.~Gong, W.~Liu, Z.~Su, M.~Guizani, {A Spatial-Temporal Transformer Network for City-Level Cellular Traffic Analysis and Prediction}, IEEE Transactions on Wireless Communications (2023).

\bibitem{yimeng2022prediction}
S.~Yimeng, L.~Jianhua, M.~Jian, Q.~Yaxing, Z.~Zhe, L.~Chunhui, {A Prediction Method of 5G Base Station Cell Traffic Based on Improved Transformer Model}, in: 2022 IEEE 4th International Conference on Civil Aviation Safety and Information Technology (ICCASIT), IEEE, 2022, pp. 40--45.

\bibitem{9363693}
{IEEE Standard for Information Technology--Telecommunications and Information Exchange between Systems - Local and Metropolitan Area Networks--Specific Requirements - Part 11: Wireless LAN Medium Access Control (MAC) and Physical Layer (PHY) Specifications}, {IEEE Std 802.11-2020 (Revision of IEEE Std 802.11-2016)} (2021) 1--4379\href {https://doi.org/10.1109/IEEESTD.2021.9363693} {\path{doi:10.1109/IEEESTD.2021.9363693}}.

\bibitem{kumar2008wireless}
A.~Kumar, D.~Manjunath, J.~Kuri, {Wireless networking}, Elsevier, 2008.

\bibitem{brownlee2018deep}
J.~Brownlee, {Deep learning for time series forecasting: predict the future with MLPs, CNNs and LSTMs in Python}, Machine Learning Mastery, 2018.

\bibitem{cisco2016}
{Cisco}, {Radio Resource Management}, White Paper (June 2016).

\bibitem{brownlee2017long}
J.~Brownlee, Long short-term memory networks with python: develop sequence prediction models with deep learning, Machine Learning Mastery, 2017.

\bibitem{woodward2022time}
W.~A. Woodward, B.~P. Sadler, S.~Robertson, {Time series for data science: Analysis and forecasting}, CRC Press, 2022.

\bibitem{zhang2019toward}
W.~Zhang, Z.~Zhang, H.-C. Chao, M.~Guizani, {Toward intelligent network optimization in wireless networking: An auto-learning framework}, IEEE Wireless Communications 26~(3) (2019) 76--82.

\bibitem{yegnanarayana2009artificial}
B.~Yegnanarayana, {Artificial neural networks}, PHI Learning Pvt. Ltd., 2009.

\bibitem{vaswani2017attention}
A.~Vaswani, N.~Shazeer, N.~Parmar, J.~Uszkoreit, L.~Jones, A.~N. Gomez, {\L}.~Kaiser, I.~Polosukhin, {Attention is all you need}, Advances in neural information processing systems 30 (2017).

\bibitem{wu2020deep}
N.~Wu, B.~Green, X.~Ben, S.~O'Banion, {Deep transformer models for time series forecasting: The influenza prevalence case}, arXiv preprint arXiv:2001.08317 (2020).

\bibitem{shumway2000time}
R.~H. Shumway, D.~S. Stoffer, {Time series analysis and its applications}, Vol.~3, Springer, 2000.

\bibitem{talathi2015improving}
S.~S. Talathi, A.~Vartak, {Improving performance of recurrent neural network with relu nonlinearity}, arXiv preprint arXiv:1511.03771 (2015).

\bibitem{kong2017short}
W.~Kong, Z.~Y. Dong, Y.~Jia, D.~J. Hill, Y.~Xu, Y.~Zhang, {Short-term residential load forecasting based on LSTM recurrent neural network}, IEEE transactions on smart grid 10~(1) (2017) 841--851.

\bibitem{ravanelli2018light}
M.~Ravanelli, P.~Brakel, M.~Omologo, Y.~Bengio, {Light gated recurrent units for speech recognition}, IEEE Transactions on Emerging Topics in Computational Intelligence 2~(2) (2018) 92--102.

\bibitem{borovykh2017conditional}
A.~Borovykh, S.~Bohte, C.~W. Oosterlee, {Conditional time series forecasting with convolutional neural networks}, arXiv preprint arXiv:1703.04691 (2017).

\bibitem{donahue2015long}
J.~Donahue, L.~Anne~Hendricks, S.~Guadarrama, M.~Rohrbach, S.~Venugopalan, K.~Saenko, T.~Darrell, {Long-term recurrent convolutional networks for visual recognition and description}, in: Proceedings of the IEEE conference on computer vision and pattern recognition, 2015, pp. 2625--2634.

\bibitem{zhou2021informer}
H.~Zhou, S.~Zhang, J.~Peng, S.~Zhang, J.~Li, H.~Xiong, W.~Zhang, {Informer: Beyond efficient transformer for long sequence time-series forecasting}, in: Proceedings of the AAAI conference on artificial intelligence, Vol.~35, 2021, pp. 11106--11115.

\bibitem{infrastructure3}
C.~P. Infrastructure, {3.5 Administrator Guide}.

\bibitem{brownlee2017introduction}
J.~Brownlee, Introduction to time series forecasting with python: how to prepare data and develop models to predict the future, Machine Learning Mastery, 2017.

\end{thebibliography}





\end{document}